\documentclass[oldversion, preprint]{aa}  
\usepackage{graphicx}
\usepackage{txfonts}
\usepackage{longtable}
\usepackage{multirow}

\begin{document}

   \title{Electron cooling and the connection between expansion and 
flux-density evolution in radio supernovae}

\titlerunning{Electron cooling in radio supernovae}

   \author{I. Mart\'i-Vidal\inst{1}
          \and
          M. A. P\'erez-Torres\inst{2}
          \and
          A. Brunthaler\inst{1}
}

   \institute{
             Max-Planck-Institut f\"ur Radioastronomie,
             Auf dem H\"ugel 69, D-53121 Bonn (Germany) 
             \email{imartiv@mpifr-bonn.mpg.de}
         \and
             Instituto de Astrof\'isica de Andaluc\'ia (CSIC)
             Apdo. Correos 2004, 08071 Granada (Spain)
}

   \date{Accepted for publication in A\&A}
 
  \abstract
{
Radio supernovae (RSNe) are weak and rare events. Their typical maximum 
radio luminosities are of the order of only $10^{27}$\,erg\,s$^{-1}$\,Hz$^{-1}$. 
There are, however, very few cases of relatively bright (and/or close) RSNe, 
from which the expansion of the shock 
and the radio light curves at several frequencies have been monitored covering 
several years. Applying the standard model of radio emission from supernovae, 
it is possible to relate the defining parameters of the modelled expansion curve 
to those of the modelled light curves in a simple algebraic way, 
by assuming an evolution law for the magnetic field and for the energy density
of the population of synchrotron-emitting electrons. However,
cooling mechanisms of the electrons may
affect considerably this connection between light curves and expansion
curve, and lead to wrong conclusions on the details of the electron 
acceleration and/or on the CSM radial density profile. In this paper, we study 
how electron cooling modifies the flux-density decay rate of RSNe 
for a set of plausible/realistic values of the magnetic field and for different 
expansion regimes. We use these results to estimate the magnetic fields of different 
RSNe observed to date and compare them to those obtained by assuming energy 
equipartition between particles and magnetic fields. For some of the best monitored 
RSNe, for which deceleration measurements, optically thin spectral index, and 
power-law time decay have been observed (SN\,1979C, SN\,1986J, SN\,1993J, and 
SN\,2008iz), we find self-consistent solutions for the index of the power-law
circumstellar density profile ($s=2$ for all cases), the index of 
the power-law relativistic electron population (rather steep values, 
$ p = 2.3 - 3.0$) and the initial magnetic field (ranging from $\sim 20$ to 
$> 100$\,G).}

   \maketitle

\section{Introduction}
\label{I}

Radio supernovae (RSNe), which are the radio counterparts of core-collapse 
supernovae (SNe), are weak and rare events. Only about $10-20$\% of the observed 
SNe are detected in radio (e.g., Weiler et al. \cite{Weiler2002}). Moreover, 
their typical maximum radio luminosities are of the order of 
$10^{27}$\,erg\,s$^{-1}$\,Hz$^{-1}$ (flux densities of the order of 1\,mJy 
for extragalactic distances, close 
to the sensitivity limits of present detectors). There are, however, very few 
cases of relatively bright RSNe, from which the expansion curve of the shock,
using Very Long Baseline Interferometry (VLBI) observations, 
and radio light curves at several frequencies were obtained covering, in some 
cases, several years, e.g.: SN\,1979C, SN\,1986J, SN\,1993J, and SN\,2008iz. 
Although there are only a handful of objects, their detailed study allowed to 
check and refine the current theoretical models of radio emission in 
supernovae. This small number of well-observed RSNe may also dramatically 
increase in the near future, thanks to the forthcoming ultra-sensitive 
interferometers with a high spatial resolution, like the Square Kilometre 
Array (SKA). 

Using the standard model of radio emission from supernovae (Chevalier 
\cite{Chevalier1982a},\cite{Chevalier1982b}), 
it is possible to relate the defining parameters of the modelled expansion curve 
to those of the modelled light curves in a simple algebraic way, 
by assuming an evolution law for the magnetic field (and for the density of the 
population of synchrotron-emitting electrons) and a radial density profile for the 
circumstellar medium, CSM, (see, e.g. Weiler et al. \cite{Weiler2002}). The 
decay in the radio-light curves according to this model is related to the time 
decay in the magnetic field and the radial decay of CSM density.
However, the continuous energy loss by the relativistic electrons (electron 
cooling), mainly due to 
synchrotron radiation (i.e., radiative cooling), but also to adiabatic 
expansion and inverse-Compton scattering, are not considered in the 
derivation of this relationship between light curves and expansion curve.
Electron cooling 
may affect considerably the shape of the light curves 
for a given expansion curve. For instance, Mart\'i-Vidal et al. (\cite{MartiVidalII}) 
succesfully modelled the exponential-like decay of the SN\,1993J radio light 
curves at late epochs, reported in Weiler et al. (\cite{Weiler2007}), 
using {\em only} radiative-cooling effects, and assuming that the density of the 
CSM was negligibly small at large distances to the progenitor 
star\footnote{An additional effect due to the escaping of the electrons from 
the emitting region might also be necessary to model the light curves of SN\,1993J, 
were the density of the CSM not negligible at those large distances to the progenitor.}.
In any case, it seems clear that if electron cooling is not considered in the 
modelling of the radio light curves of a supernova, it could result into wrong 
estimates of the model parameters.
In this paper, we study how electron
cooling modifies the flux-density decay rate of RSNe for several values 
of the magnetic field and for different expansion regimes. These results 
can be used to estimate the magnetic fields of observed RSNe. 

In the next 
section, we outline 
the standard model of radio emission from supernovae. In Sect. \ref{III} 
we study the effect of electron cooling in the population of emitting 
electrons and in the flux-density decay rate. In Sect. \ref{IV} we present 
the results of several simulations of the expansion and radio light curves 
of RSNe. In Sect. \ref{V}, we explain how these results can be used in real 
cases to estimate physical quantities in RSNe and estimate the magnetic fields 
for several observed RSNe, comparing these 
estimates to those obtained by assuming particle-field energy 
equipartition. 
In Sect. \ref{VI} we 
summarize our conclusions.

\section{Connection between expansion and radio light curves in RSNe}
\label{II}

In the standard model of emission from supernovae (Chevalier 
\cite{Chevalier1982a}, \cite{Chevalier1982b}), the spherically-symmetric 
expanding shock is described as a contact discontinuity plus two shocks, one
moving backwards (from a Lagrangian point of view) and the other moving forward,
shocking the CSM. A fraction of shocked CSM electrons is accelerated to 
relativistic energies, possibly due to statistical Fermi processes, and produce 
synchrotron emission at radio wavelengths as they interact with high 
magnetic fields in the shocked CSM region.

The distance, $r$, from the contact discontinuity to the center of the expansion
evolves as a power-law of time ($r \propto t^m$) with an {\em expansion index}, 
$m$, that depends on the radial density profiles of CSM 
($\rho_{CSM} \propto r^{-s}$) and ejecta ($\rho_{ej} \propto r^{-n}$) in the 
form (Chevalier \cite{Chevalier1982a}, \cite{Chevalier1982b})

\begin{equation}
m = \frac{n-3}{n-s}.
\label{mEq}
\end{equation}

This solution of the shock 
expansion holds for $n>5$ and $s<3$. The structure of the shock (contact
discontinuity plus backward and forward shocks) expands in a self-similar way. 
Therefore, the expansion of the forward and backward shocks also follows the 
law $\propto t^m$.

On the other hand, the distribution of relativistic electrons in energy space 
follows a power law ($N \propto E^{-p}$) and the 
energy-density of the magnetic field is assumed to be proportional to the 
energy-density of the shock (i.e., $B^2 \propto n\,V^2$, where $B^2$ is the 
average magnetic field squared, $n \propto r^{-s}$ is the particle number density, 
and $V \propto r^{(m-1)/m}$ is the shock expansion velocity). Hence, 

\begin{equation}
B \propto t^{m\,(2-s)/2-1}.
\label{magEq}
\end{equation}

We must notice a limitation in the standard model at this point. For a standard
CSM particle density of $10^8$\,cm$^{-3}$ at a distance of $10^{15}$\,cm from the explosion
center, and an expansion velocity of $20\,000$\,km\,s$^{-1}$, a magnetic field of 
50--60\,G translates into a similar energy density for the expanding shock and the magnetic
field. Such a large magnetic-field energy density may affect the hydrodynamics of the 
shock\footnote{Detailed magneto-hydrodynamic simulations would be necessary to study 
the real impact of large magnetic fields in the evolution of the expanding shock}.
This effect is neglected in the model (which, indeed, assumes that the magnetic-field 
energy density is a small fraction of that of the shock). Hence, for cases of very large
magnetic fields reported in Sect. 5, high CSM particle densities and/or large 
expansion velocities might be accordingly considered, to make the magnetic-field estimates
consistent in the frame of the standard model.

The fraction of accelerated particles by the shock, or {\em injection efficiency}  
of the shock, is also assumed to be proportional to the shock energy density. 
Under all these assumptions, and considering that the intensity of synchrotron radiation 
is (e.g. Pacholczyk \cite{Pacholczyk1970})   

\begin{equation}
I \propto N\,B^{(1+p)/2},
\label{IBN}
\end{equation}

\noindent it is possible to derive the intensity, $I$, in the optically-thin regime
if we neglect electron cooling. Since, in that case, 
$\dot{N}(E) \propto E^{-p}\,n\,r^2\,V\,\mathrm{d}t$, it can be shown that 
$I \propto \nu^{-\alpha}\,t^{\beta}$, with

\begin{equation} 
\alpha = \frac{p-1}{2}.
\label{SpecIndEq}
\end{equation}

\noindent and

\begin{equation}
\beta = \frac{1}{4}(m\,(2\,(11 + p) - (5 + p)\,s) - 2p - 10).
\label{betaEq}
\end{equation}

This equation brings a direct relation between the decay index of the radio 
light curves in their optically-thin regime, $\beta$, on one hand, and the supernova 
expansion index, $m$, the energy index of the injected relativistic electrons, $p$, and 
the index of the CSM radial density profile, $s$, on the other hand. For the case of a 
constant pre-supernova mass-loss wind (i.e., $s=2$) this equation
reduces to $\beta = (6m-p-5)/2$ (e.g., Weiler et al. \cite{Weiler2002}).

\section{Radiative and adiabatic cooling of the relativistic electrons}
\label{III}

The supernova shock is continuously accelerating electrons from the shocked
CSM. These electrons are distributed as $N \propto E^{-p}$. However, the electrons 
already shocked that are emitting synchrotron radiation loose energy and, therefore, 
shift towards lower energies in the electron-energy distribution. Since the number of 
electrons is conserved, we can make use of the continuity equation in energy space, i.e.,

\begin{equation}
\dot{N} = \nabla_E(N\dot{E}) + S(E,t) - L(E,t),
\label{contEq}
\end{equation}

\noindent where $S(E,t)$ is the source function (the new electrons continuously 
accelerated by the shock) and $L(E,t)$ accounts for the escaping of electrons from the emitting
region. We will assume that $L(E,t) = 0$ (in Mart\'i-Vidal et al. \cite{MartiVidalII} 
we use $L(E,t) \propto N$ to model the SN\,1993J radio data, although the effects of this 
term are very small compared to $S(E,t)$ until very late epochs, when a large drop in the 
CSM density profile takes place). It can be shown (see Appendix \ref{SourceApp}) that the 
source function is $S(t)\,E^{-p}$, where

\begin{equation}
S(t) = N_0\,F_{rel}\,\frac{p-1}{E_{m}^{1-p}}\,\left(\frac{t}{t_0}\right)^{m(5-s)-3},
\label{SourceEq}
\end{equation}

\noindent where $N_0$ is the number density of shocked CSM electrons at a reference 
epoch ($t_0$), $F_{rel}$ is the fraction of accelerated electrons (of the order of 
$10^{-5}$ for SN\,1993J), and $E_m$ is the minimum energy of the relativistic electrons 
(we set $E_m = m_e \,c^2$, although this value is not relevant in the optically-thin
regime of the light curves). 

The term $\dot{E}$ takes into account the energy loss (or gain) of the electrons. The 
energy loss can be either radiative, adiabatic, and/or due to free-free interactions with atoms 
or ions in the CSM. The energy gain can be due to self-absorption of the synchrotron 
radiation or to inverse-Compton scattering, although these effects are negligible in the 
optically-thin part of the light curve (and also for large magnetic fields), which 
is that of our interest here. In the case of radiative losses, we have

\begin{equation}
\dot{E}_{\mathrm{r}} = - c_2 B_\perp^2\,E^2\,\left(\frac{t}{t_0}\right)^{m(2-s)-2},
\label{DotEEq1}
\end{equation}

\noindent where $c_2 = 2.37\times10^{-3}$ in cgs units (see Pacholczyk \cite{Pacholczyk1970}), 
and $B_\perp$ is the magnetic field at a reference epoch ($t_0$) averaged in the orthogonal 
planes to the electron trajectories. For a random distribution of magnetic-field lines and 
electron trajectories, $B_\perp$ is equal to $\sqrt{2/3}$ times the total averaged magnetic 
field, $B_0$, at the reference epoch.
In the case of adiabatic losses, we have

\begin{equation}
\dot{E}_{\mathrm{a}} = \frac{1}{r}\frac{d\,r}{d\,t}\,E = m\frac{E}{t}.
\label{DotEEq2}
\end{equation}

Therefore, if radiative cooling and adiabatic expansion are the dominant processes of 
energy loss by the electrons, we have

\begin{equation}
\dot{E} = - c_2 B_\perp^2\,E^2\,\left(\frac{t}{t_0}\right)^{m(2-s)-2} -m\frac{E}{t},
\label{DotEEq}
\end{equation}

In Eq. \ref{DotEEq}, we have neglected the term due to free-free interactions of the 
electrons with the surrounding CSM atoms and ions
($\dot{E} \propto r^{-s}\,E$), since this term is much smaller than the radiative and 
adiabatic terms in the optically-thin regime of the light curves. 
In Appendix \ref{RadAdiFreeApp}, we analyze under which 
conditions might the free-free term not be negligible compared to the radiative and 
adiabatic terms. 

Equation \ref{contEq}, together with Eqs. \ref{SourceEq} and \ref{DotEEq}, is a typical 
difussion-like partial differential equation that can be numerically integrated using, for 
instance, a semi-implicit approach (e.g. Mart\'i-Vidal et 
al. \cite{MartiVidalII}). However, since synchrotron self-absorption, inverse Compton, 
and free-free interactions are neglected (i.e., only the radiative and adiabatic terms in 
$\dot{E}$ are considered), it is also possible to find an integral form for the 
solution of this simplified version of Eq. \ref{contEq}. We show this solution in 
Appendix \ref{AppA}.
From the numerical solution of $N(E,t)$, we can estimate the flux-density 
decay rate of the light curves, since the intensity is 

$$ I \propto t^{m(2-s)/2-1}\,\int_{E_m}^{\infty}{N\,F(x)\,dE},$$

\noindent where the power-law of time is related to the decay of the magnetic 
field (see Eq. \ref{magEq}), $x$ is the ratio between the observing frequency and the critical 
frequency at energy $E$, and $F(x)$ is

$$ F(x) = x \int_{x}^{\infty}{K_{5/3}(z)\,dz},$$

\noindent being $K_{5/3}(z)$ a Bessel function of the second kind 
(e.g., Pacholczyk \cite{Pacholczyk1970}). Then, from the time evolution of $I$, we 
can estimate $\beta$ for different combinations of $m$, $p$, $s$, $B_0$,
$N_0$, and $F_{rel}$, and compare the results to Eq. \ref{betaEq} in order to 
check the effect of electron cooling in the light curves.

\section{Effect of magnetic fields in the radio light curves}
\label{IV}

\begin{figure*}[t!]
\centering
\includegraphics[width=17cm]{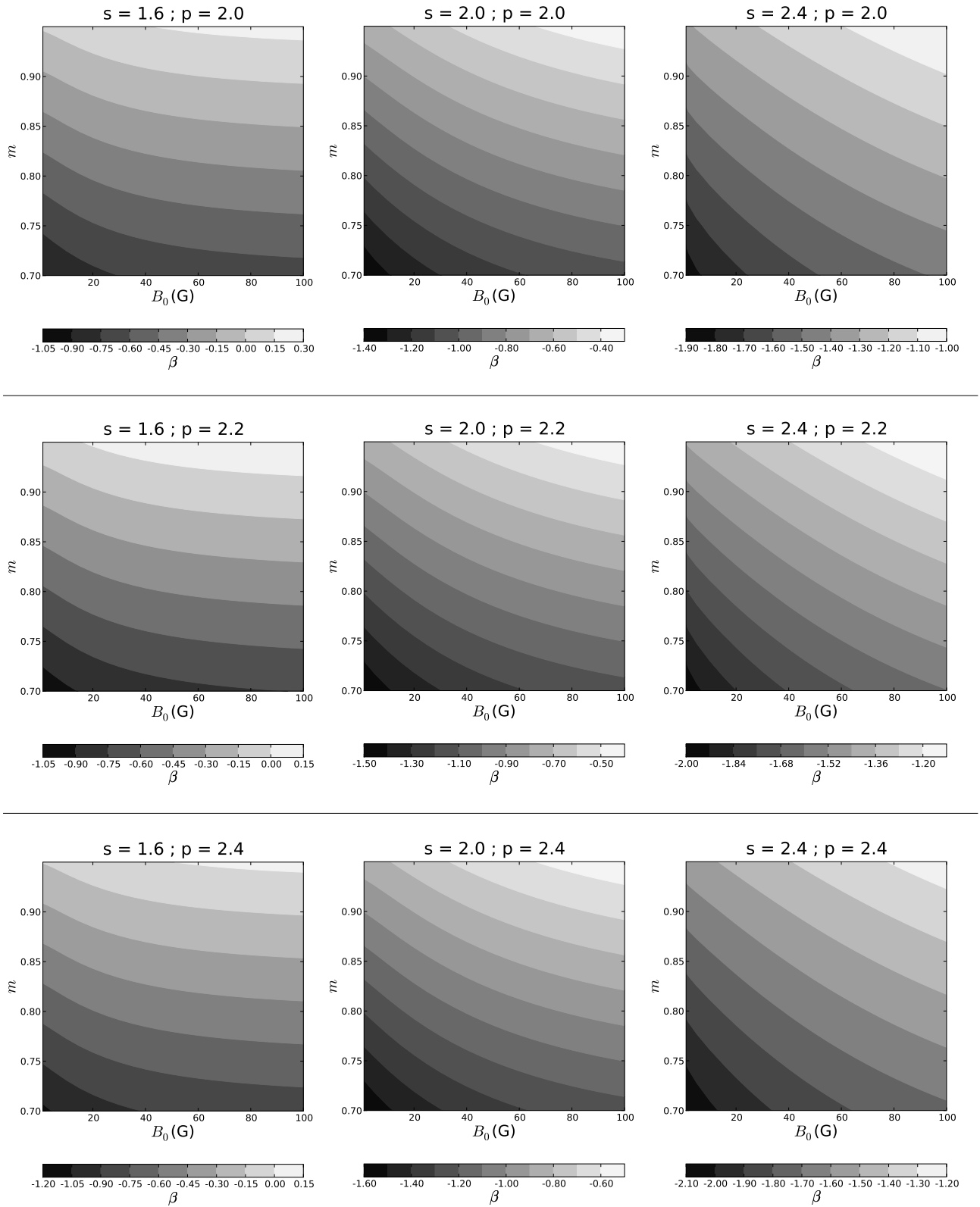}
\caption{$\beta$ as a function of 
$m$ and $B_0$ (reference epoch $t_0 = 5$\,days) for a selection of values for $s$ and $p$.
For $B_0 \sim 0$, we roughly recover the $\beta$ given in Eq. \ref{betaEq}. As we increase 
$B_0$, $\beta$ approaches 0. This effect is more pronounced for larger values of $s$ (i.e., 
for steeper CSM radial density profiles).}
\label{fig1}
\end{figure*}

\begin{figure*}[t!]
\centering
\includegraphics[width=17cm]{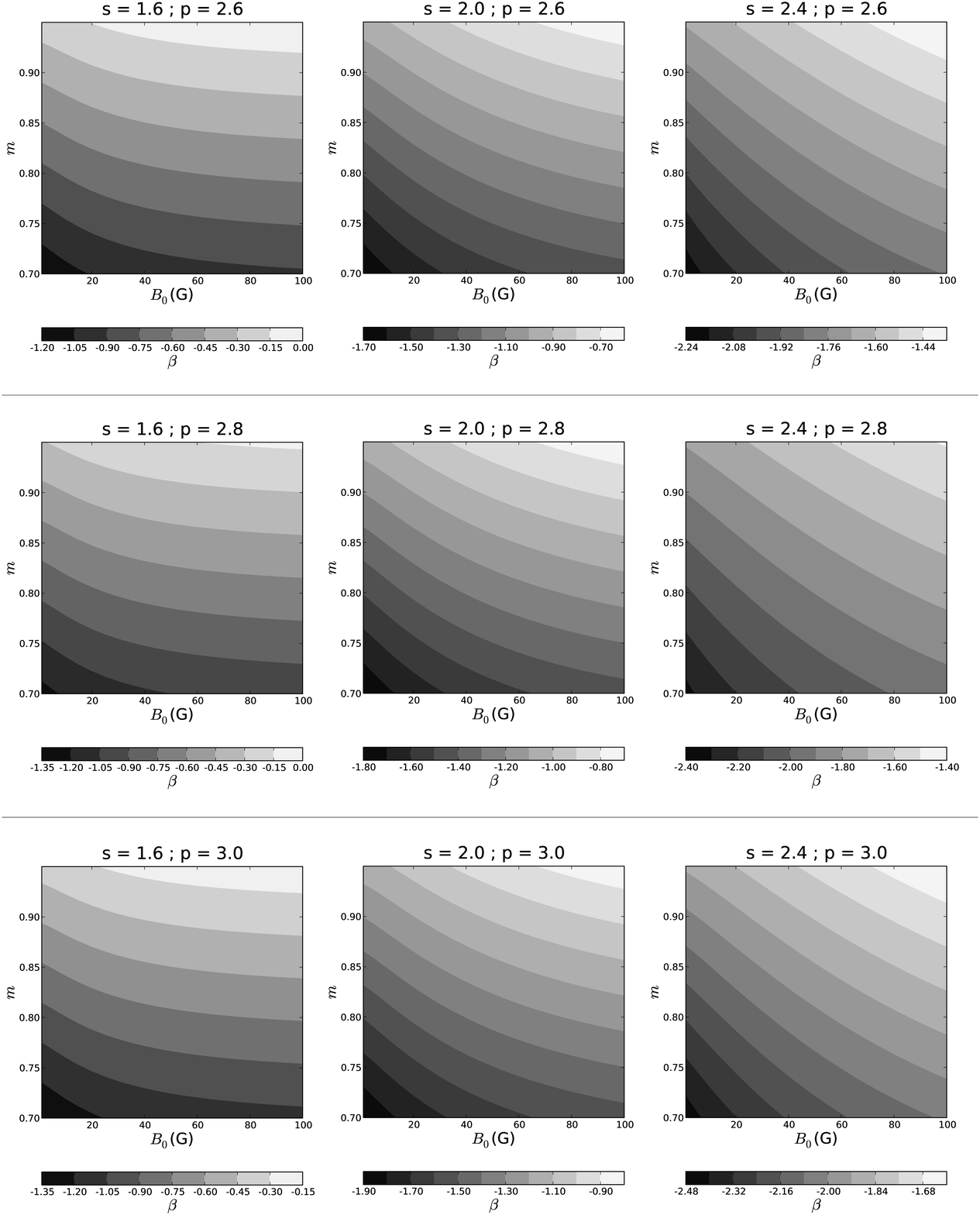}
\caption{Same as Fig. \ref{fig1}, but for a different set of values for $p$.}
\label{fig1b}
\end{figure*}

We show in Figs. \ref{fig1} and \ref{fig1b} the $\beta$ obtained from our simulations 
as a function of $m$ and $B_0$ (the magnetic field at the reference epoch $t_0 = 5$\,days) 
for 6 values of $p$ 
(2.0, 2.2, and 2.4, in Fig. \ref{fig1}; 2.6, 2.8, and 3.0, in Fig. \ref{fig1b}) 
and for 3 values of $s$ (1.6, 2.0, and 2.4).
We have computed $\beta$ at 5\,GHz between 300 and 1000 days after the shock breakout. 
Different selections of frequencies and/or age ranges result into deviations 
in $\beta$ of a few \% at most.

Since $\beta$ is computed in the optically-thin part of the radio light curves,
$N_0$ is not really important in the simulations (changing this value would affect
the opacity in the early supernova evolution). In our case, the important quantity
would be $N_0\,F_{rel}$, which accounts for the number of {\em relativistic} electrons.
Indeed, $N(E,t)$ only depends on $B_0$ regardless of a constant
scaling factor defined by $N_0\,F_{rel}$. Therefore, the value of $N_0\,F_{rel}$
does not really affect the estimates of $\beta$.
To ensure that this statement is correct, we checked that the values of $\beta$
derived from our simulations are only sensitive to changes in
$B_0$, $m$, $p$, and $s$.

If radiative cooling is negligible (i.e., for small values of $B_0$), the $\beta$ 
computed from our simulations approaches the values computed from Eq. \ref{betaEq} for 
all combinations of $m$, $p$, and $s$. This is an expected result, since the adiabatic 
losses alone (which are $\propto E$) do not affect the power law of the electron 
distribution (Pacholczyk \cite{Pacholczyk1970}). However, as the magnetic field 
increases, $\beta$ decreases in absolute value (i.e., the light curves become flatter).
This result is in principle non-intuitive, since one would expect the light curves
to be steeper as the radiative cooling (i.e., the energy loss of the electrons) 
is more important. The light curves in the optically-thin stage are flatter for 
larger $B_0$, because the magnetic field decreases as the supernova expands (see Eq. 
\ref{magEq}) and,
therefore, cooling effects (which are smaller for smaller magnetic fields) are less
important as time goes by. Thus, $\dot{N}$ assymptotically approaches the value
without cooling as the supernova expands. As a consequence, $\dot{N}/N$ (which affects
the value of $\beta$) takes a larger value if we consider radiative cooling. In Appendix
\ref{betaradApp}, we show the details of this discussion mathematically.

The largest deviations of $\beta$ with respect to the cooling-free 
value (i.e., that of Eq. \ref{betaEq}) correspond, in all cases, to the smallest
decelerations of the shock (i.e., values of $m$ close to 1) and/or to the 
steepest CSM radial density profiles (i.e., larger values of $s$).

In the case $s=2$, we can approximate the $\beta$ shown in Figs. \ref{fig1} and 
\ref{fig1b} with the phenomenological equation

\begin{equation}
\frac{\beta(B_0)}{\beta(B_0=0)} = \left(\frac{F_1/(1-m)}{B_0+F_1/(1-m)}\right)^{F_2/((1-m)(2p-1))},
\label{ApproxBetaEq}
\end{equation}

\noindent where $\beta(B_0)$ corresponds to a magnetic field $B_0$ and $\beta(B_0=0)$ is that
given in Eq. \ref{betaEq} (i.e., with no radiative cooling considered). The parameters $F_1$ 
and $F_2$ take the values 7.725\,G and 0.184, respectively. The maximum deviation between 
the $\beta$ computed from Eq. \ref{ApproxBetaEq} and those shown in Figs. \ref{fig1} and 
\ref{fig1b} (for $s=2$) is only 3.5\%.

\subsection{Changes in the spectral index}

\begin{figure*}[t!]
\centering
\includegraphics[width=17cm]{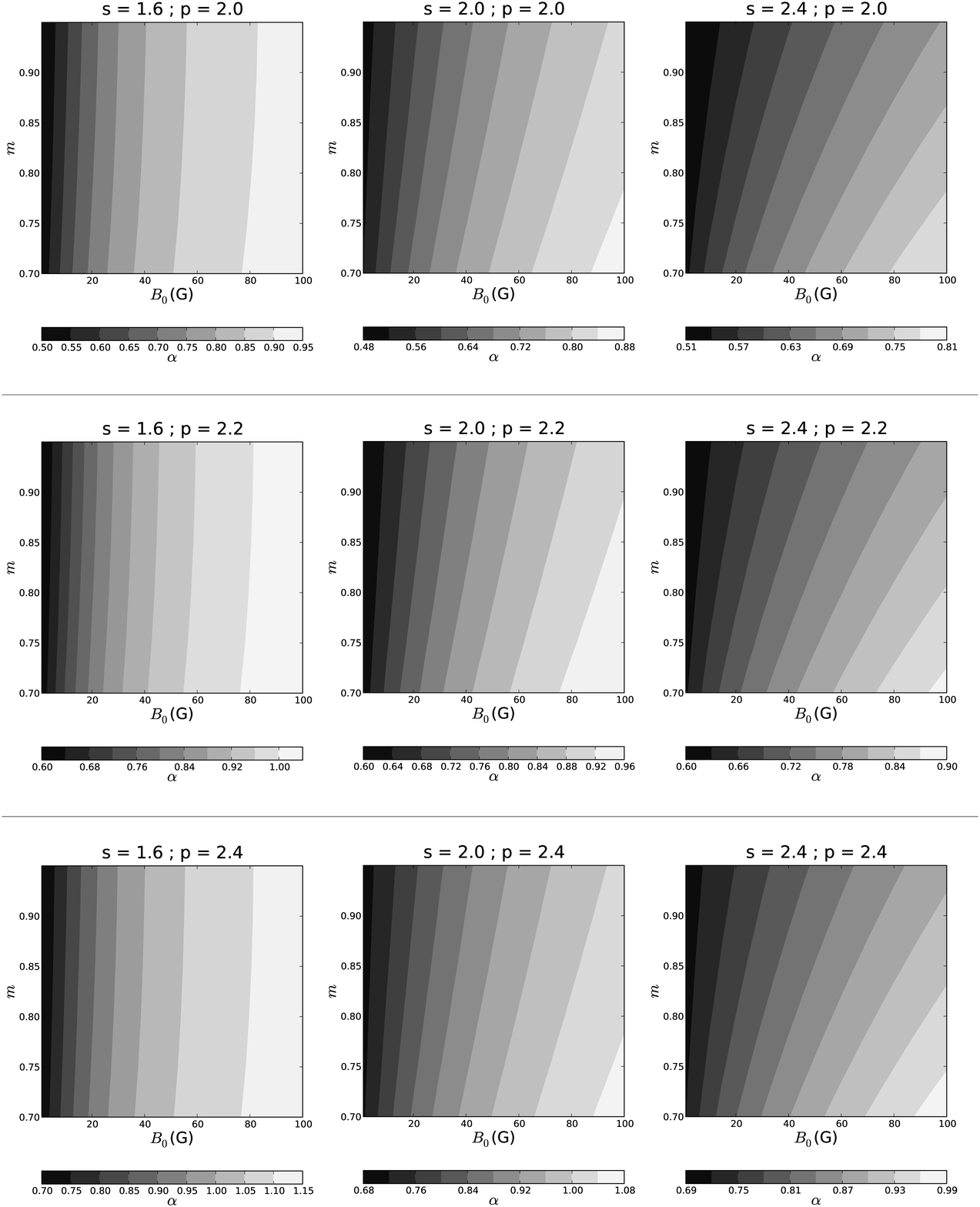}
\caption{Spectral index $\alpha$, centered at 5\,GHz, as a function of $m$ and 
$B_0$ (reference epoch 
$t_0 = 5$\,days) for a selection of values for $s$ and $p$. For $B_0 \sim 0$ we 
roughly obtain the canonical value $\alpha = (p-1)/2$. As we increase $B_0$, the 
spectra become steeper ($\alpha$ increases). This effect is more pronounced for 
lower values of $s$ (i.e., for flatter CSM radial density profiles).}
\label{fig2}
\end{figure*}

\begin{figure*}[t!]
\centering
\includegraphics[width=17cm]{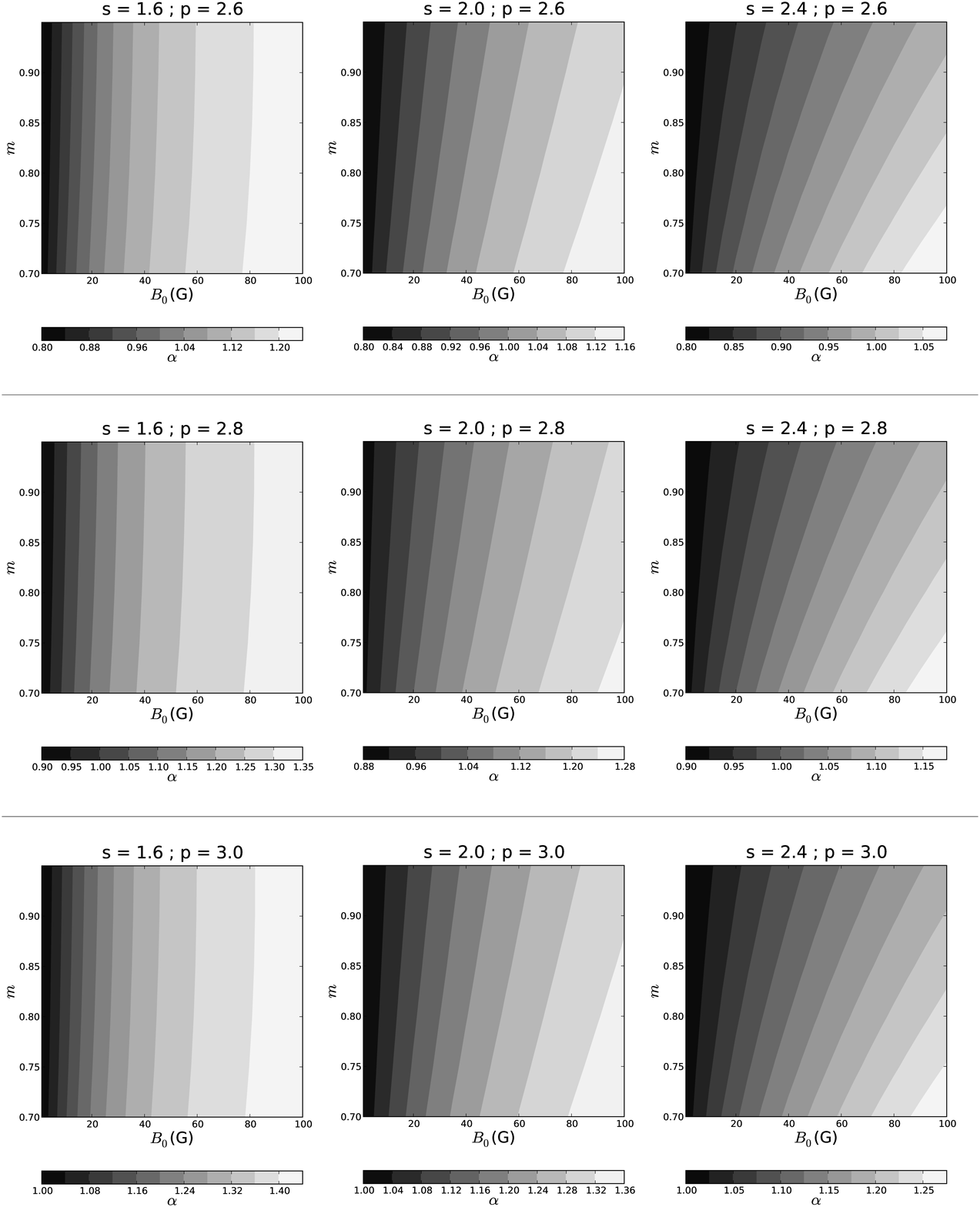}
\caption{Same as in Fig. \ref{fig2}, but for a different set of values for $p$.}
\label{fig2b}
\end{figure*}

If electron cooling is not considered, there is a direct relationship between $p$ 
and the spectral index $\alpha$ (see, Eq. \ref{SpecIndEq}): $p = 1 + 2\alpha$.
However, when electron cooling is taken into account, there is a flux of 
electrons towards smaller $E$, which increases the value of $\alpha$. This effect 
is more important as we increase the observing frequency. We must notice, 
however, that new electrons are continuously being injected in the emitting region, 
and their energy distribution is assumed to be always $\propto E^{-p}$, so this fraction
of electrons is not affected by cooling. Therefore, 
the effect of cooling in the spectral-index steepening is somewhat mimicked by 
the new electrons entering the shocked CSM. 
The integration of Eq. \ref{contEq} takes into
account this trade-off between electron cooling and the source function. In Figs. 
\ref{fig2} and \ref{fig2b}, we show the simulated spectral indices, averaged between 
300 and 1000 days after shock breakout and centered at 5\,GHz. We show $\alpha$ for the 
same values of $p$ and $s$ used in Figs. \ref{fig1} and \ref{fig1b}. 

We notice that radiative cooling is more important at higher energies, so the 
(effective) spectral index should slightly increase with the observing frequency. 
for instance, the difference between the spectral indices centered at 5\,GHz (which
are higher) and those at 1.7\,GHz (which are lower) is typically 2--3\% for 
magnetic fields of 10\,G and 5--6\% for magnetic fields of 100\,G.

The values of $\alpha$ obtained from our simulations tend to the expected 
values without cooling (i.e. $\alpha = (p-1)/2$) when the magnetic field approaches
0, also as expected. An increase in $B_0$ steepens the spectrum of the radiation 
(i.e., $\alpha$ increases), for all combinations of $s$, $m$, and $p$, because 
$\dot{E}_{\mathrm{rad}} \propto E^2$.

In the case $s=2$, we can also approximate the $\alpha$ shown in Figs. \ref{fig2} and
\ref{fig2b} with a phenomenological equation

\begin{equation}
\frac{\alpha(B_0)}{\alpha(B_0=0)} = \left(\frac{B_0}{F_1(2m-1)(2p+5)}+1\right)^{m/(2p-1)},
\label{ApproxAlphaEq}
\end{equation}

\noindent where $\alpha(B_0)$ corresponds to a magnetic field $B_0$ and $\alpha(B_0=0)$ is $(p-1)/2$ 
(i.e., no radiative cooling considered). The parameter $F_1$ takes the value 3.04\,G, and the 
maximum deviation between the $\alpha$ computed from Eq. \ref{ApproxAlphaEq} and those from the 
simulations is also 3.5\%, for all the analyzed values of $m$, $p$, and $B_0$.

Figures \ref{fig1}, \ref{fig1b}, \ref{fig2}, and \ref{fig2b} (and eventually Eqs. \ref{ApproxBetaEq} 
and \ref{ApproxAlphaEq}) can be used to estimate the
magnetic field in a supernova by using the $\alpha$, $m$, and $\beta$ inferred from the
observations (provided light curves and the expansion curve of the supernova have been
observed). In the next section, we will estimate magnetic
fields in a number of radio supernovae, for which expansion curve and
radio light curves are available.

\subsection{The special case $p=2$}
\label{p2sec}

We notice that for the special case $p=2$, the effect of radiative cooling in the 
electron energy distribution should be negligible for all $E$, since $N\dot{E}_{\mathrm{r}}$ 
would not depend on $E$ and its contribution to the energy gradient of $N$ would therefore be 
null (see Eq. \ref{contEq}). 
In principle, one would expect the population of electrons to evolve as if there were only 
adiabatic cooling, so neither $\beta$ nor $\alpha$ should depend on the magnetic field. However,
when $p=2$, the evolution of $N(E,t)$ is not only determined by the source function, $S(E,t)$, 
and the adiabatic term, but also by $t_F$ (see Appendix \ref{AppA}, Eq. \ref{ExactSol}), which 
is the time at which all the electrons with energies larger than $E$ have energies below $E$ at 
time $t$. The time $t_F$ is larger than $t_0$ for high energies and/or large $t$, and 
depends on the magnetic field. Thus, even for $p=2$, the light curves and spectra will be 
somewhat modified by radiative electron cooling at high frequencies and late epochs (those 
frequencies and supernova ages depend, of course, on the strength of the magnetic field and 
the deceleration of the shock), as it is shown in Figs. \ref{fig1} and \ref{fig2} (upper rows).

\section{Estimate of magnetic fields in observed RSNe}
\label{V}

If a radio supernova is strong enough to be monitored with VLBI, it is possible
to estimate $m$ from the expansion curve and $\beta$ and $\alpha$ from the light
curves\footnote{We assume that all these quantities are determined in the 
optically-thin regime, which corresponds to a positive $\alpha$ (i.e., a steep 
spectrum) and a decreasing (or non-increasing) flux density, with the exception
of very special cases ($s<<2$ together with $m\sim1$, see Figs. 1 and 2). In all the
observational cases studied in this paper, the conditions for an optically-thin regime 
hold for all the quantities used in our analysis.}. 
If cooling is not considered, from $\alpha$ it is possible to derive 
$p$ and, using Eq. \ref{betaEq}, it is possible to derive $s$. Additionally, 
assuming a constant temperature of the CSM electrons, the opacity due to 
free-free processes decreases as $t^{\delta}$, being $\delta = m\,(1-2s)$ (e.g., 
Weiler et al. \cite{Weiler2002}). Therefore, if the light curves are well sampled in the 
optically-thick regime, another condition can be imposed to the parameters 
if we assume dominance of free-free absorption (FFA) over synchrotron self 
absorption (SSA). Self-consistency between all the parameters can then be checked.

However, it is not clear how much FFA dominates the light curves of usual RSNe. 
For instance, SSA has shown to be, by far, the dominant absorption mechanism in 
all the evolution of the SN\,1993J light curves (Fransson \& Bj\"ornsson 
\cite{Fransson1998}; P\'erez-Torres et al. \cite{PerezTorres2002}; 
Mart\'i-Vidal et al. \cite{MartiVidalII}). Moreover, different forms of electrons 
cooling, as we have shown in the previous section, affect the values
of $\beta$ and $\alpha$ for a set of $m$, $s$, and $p$, depending on the strength of the 
amplified magnetic field. In this section, we will infer the values of magnetic fields of 
several RSNe, based on the the values of $\alpha$, $\beta$ and $m$ estimated from the 
observations. An a priori value for $s$ and/or $p$ must be however assumed to estimate 
$B_0$ using this approach.

\subsection{SN\,1979C}

Weiler et al. (\cite{Weiler1991}) reported more than 10 years of observations 
of the SN\,1979C radio light curves at 15, 5, and 1.4\,GHz. These authors fit 
$\alpha = 0.74^{+0.05}_{-0.08}$ and $\beta = -0.78^{+0.02}_{-0.03}$. In regard 
of the expansion curve, 
different results have been reported by different authors. Marcaide et al. 
(\cite{Marcaide2002}) reported a strong deceleration in SN\,1979C and, 
from a more complete VLBI dataset, Bartel \& Bietenholz (\cite{Bartel2003}) 
reported an almost free expansion (i.e., $m \sim 1$) for 22 years. More 
recently, Marcaide et al. (\cite{Marcaide2009b}) re-analyzed their VLBI data 
and complemented them with new 1.6\,GHz observations and the data from Bartel 
et al. These authors arrived to the conclusion that the expansion of SN\,1979C 
was indeed almost free ($m = 0.91 \pm 0.09$) for over 25 years.  

The fitted $\alpha$ is very close to the value corresponding to $p=2.5$ without
radiative cooling. Therefore, we conclude that either the 
magnetic field is very small (and hence $\alpha \sim (p-1)/2$), or $p$ is lower 
than 2.5. Assuming $p=2.2$ (or $p=2.4$) and $s=2$, we estimate from Fig. \ref{fig1} 
a magnetic field of $\sim20$\,G (or $\sim40$\,G) at day 5. There are no solutions 
neither for $s=1.6$ nor $s=2.4$. Now, from Fig. \ref{fig2}, 
the observed $\alpha$ and $m$ imply $B_0 \sim 20-30$\,G (for $p=2.2$) and 
$B_0 \sim 5-10$\,G (for $p=2.4$). Therefore, based on the radiative-cooling assumption, 
the magnetic field at day 5 should be between 20 and 30\,G if $p \sim 2.2$. 
Indeed, from Eqs. \ref{ApproxAlphaEq} and \ref{ApproxBetaEq} we find self-consistent 
estimates of $\alpha$ and $\beta$ for $p=2.3$ and $B_0 = 30$\,G.

How do these estimates ob $B_0$ compare to the equipartition magnetic field?
In the case of energy equipartition between particles and fields, it is possible 
to estimate the magnetic field in the radiating region provided the size and the 
total luminosity of the source are known. The expression used for this estimate 
is taken from Pacholczyk (\cite{Pacholczyk1970}):

\begin{equation}
B_{eq} = (4.5 c_{12} (1 + k)/\phi)^{2/7} R^{-6/7} L^{2/7}_R ,
\label{EquipEq}
\end{equation}

\noindent where $c_{12}$ depends on the spectral index, $\alpha$, and on the 
frequency range in the spectrum integration. $\phi$ is the filling factor of 
the emitting region to a sphere (0.66 for a shell-like structure of 30\% 
fractional width), $R$ is the source radius, $L_R$ is the 
integrated radio luminosity, and $k$ is the ratio between the heavy particle 
energy density to the electron energy density. We do not know the details of 
the particle acceleration, and the efficiency of acceleration could depend 
on the particle mass. Hence, $k$ can vary from 1 (case of a much larger 
acceleration efficiency of the electrons compared to the protons) to 
$m_p/m_e \sim 2 \times 10^3$ (case of a similar acceleration efficiency 
for electrons and protons).

Using the spectral index and flux densities by Weiler et al. 
(\cite{Weiler1991}), the expansion curve by Marcaide et al. 
(\cite{Marcaide2009b}), and the distance to the host galaxy (M\,100)
by Ferrarese et al. (\cite{Ferrarese1996}) of 16.1\,Mpc, we estimate 
$L_R = 1.6\times10^{33}$\,erg\,s$^{-1}$ at day 5 after explosion. Therefore, 
the equipartition magnetic field could range between $10$ and $85$\,G (for 
$k=1$ and $k=2000$, respectively). Our estimated $B_0$, 
assuming $p=2.3$ and $s=2$, corresponds to low-to-intermediate values of $k$, 
i.e. low-to-intermediate efficiency in the acceleration of ions.

We must notice that the cooling-free prediction of $\beta$ (Eq. \ref{betaEq})
for $s=2$ and $p=2.5$ is consistent with the observed one if 
$m=0.99$ (i.e. for an essentially non-decelerated expansion), which is indeed 
compatible with the value of $m$ reported in Marcaide et al. (\cite{Marcaide2009b})
at a 1$\sigma$ level. This latter possibility would imply a very small magnetic-field
energy density, compared to the energy density of the particles.

In Table \ref{SelfConsTab} (row 1) we summarize the values of $s$, $p$, and $B_0$ 
discussed for this supernova.

\begin{table}
\begin{minipage}[t]{9cm}
\caption{Model parameters for several RSNe. {\em Observed} refers to those 
obtained from the fitted expansion and radio light curves; {\em Assumed} and 
{\em Derived} refer to those obtained from comparison with the results shown in 
Figs. \ref{fig1}, \ref{fig1b}, \ref{fig2}, and \ref{fig2b}. Cases with two 
possible solutions are given in two rows, one row for each solution.}
\renewcommand{\footnoterule}{}
\begin{tabular}{l | c c c | c c c}
\hline\hline
{\bf Supernova}  & \multicolumn{3}{c}{{\bf Observed}} & \multicolumn{2}{c}{{\bf Assumed}} & {\bf Derived} \\
           & $m$ & $\alpha$ & $\beta$ & $s$ & $p$ & $B_0$\,(G) \\
\hline
SN\,1979C  & 0.91 & 0.74 & $-0.78$ & 2.0 & 2.5 & $\sim 0$ \\
           &      &      &       & 2.0 & 2.3 & 20--30   \\
\hline
SN\,1986J  & 0.69 & 0.7  & $-1.18$ & 1.7 & 2.4 & $\sim 0$ \\
           &      &      &       & 2.0 & 2.0 & 30--50 \\
\hline
SN\,1993J\footnote{Fit to data between days 300 and 1000 after explosion (see text).}  
           & 0.87 & 0.98 & $-0.78$ & 2.0 & 2.5 & 60--80   \\
\hline
SN\,2008iz & 0.89\footnote{Derived from VLBI observations (Brunthaler et al. \cite{Brunthaler2011})} 
           & 1.08 & $-1.43$ & 2 & $\sim 3$ & $\sim 0$ \\
           &      &      &       & 2.4  & 2.6 & $\sim 100$ \\ 
\hline
\end{tabular}
\end{minipage}
\label{SelfConsTab}
\end{table}

\subsection{SN\,1986J}

There are a number of peculiarities in the radio light curves of SN\,1986J compared
to those of other RSNe. The spectral index cannot be well fitted to a single 
value from 1.4 to 23\,GHz (Weiler, Panagia \& Sramek \cite{Weiler1990}). In
the optically-thin part of the light curves, $\alpha = 0.7 \pm 0.1$ between 
5 and 15\,GHz, but $\alpha = 0.2 \pm 0.2$ between 15 and 23\,GHz. Additionally,
Bietenholz, Bartel, \& Rupen (\cite{Bieten2004}) reported the discovery of 
a compact source in the shell center of SN\,1986J with an inverted spectrum,
and interpreted this source as due to accretion onto a black hole or to a 
young pulsar nebula. 

The best-fit parameters reported in Weiler, Panagia \& Sramek (\cite{Weiler1990})
are $\alpha = 0.67^{+0.04}_{-0.08}$ and $\beta = -1.18^{+0.02}_{-0.04}$, based 
on observations up to year 1989. Bietenholz, Bartel, \& Rupen (\cite{Bietenholz2002}) reported
a much lower $\beta$ for later epochs that slightly depends on the observing 
frequency (ranging from $-2.7$ at 8.4\,GHz to $-3.5$ at 23\,GHz). In this work 
we will use the $\alpha$ and $\beta$ obtained from the early epochs 
(i.e., those up to year 1989) and between 5 and 15\,GHz.

In regard of the expansion curve, Bietenholz et al. (\cite{Bietenholz2010}) reported
$m = 0.69 \pm 0.03$, a value much lower than those of the other RSNe observed with 
VLBI ($\sim$0.8$-$0.9). 

Now, from the extrapolated size at day 5, a distance to the 
host galaxy (NGC\,891) of $8.4\pm0.5$\,Mpc (Tonry et al. \cite{Tonry2001}), and 
using $\alpha = 0.7$, we obtain an equipartition magnetic field between $14$ and 
$100$\,G ($k=1$ and $2000$, respectively) using Eq. \ref{EquipEq} for day 5 after 
explosion.

How do these estimates of $B_0$ compare to those that can be obtained with our 
approach? 
A spectral index of 0.7 can only be obtained with $p=2.4$ or lower. 
Trying with the lowest value, $p=2$, we find from Eqs. \ref{ApproxBetaEq} and
\ref{ApproxAlphaEq} (or Figs. \ref{fig1} and \ref{fig2}) self-consistent values 
of $\alpha$ and $\beta$ with $B_0 = 30-50$\,G.
Using $p=2.2$ 
and $s=2$, we estimate from Fig. \ref{fig1} a magnetic field 
of $\sim$60\,G at 5 days after explosion. Now, from Fig. \ref{fig2} and 
assuming the same values for $s$ and $p$, a magnetic field of $\sim$10\,G is estimated.
Both estimates are incompatible.
For $s=2.4$ or $s=1.6$ we can neither obtain a coherent estimate of the magnetic field;
using now $p=2.4$, the observed $\alpha$ requires, of course, $B_0 \sim 0$ and the 
observed $\beta$ can only be explained with our simulations if $s \sim 1.6$. Therefore, 
a compatibility between Figs. \ref{fig1} and \ref{fig2} is found for small values 
of the magnetic field and a rather flat CSM radial density profile ($s \sim 1.6$).
Indeed, the cooling-free prediction of $\beta$ given by Eq. \ref{betaEq}
(which is similar to that one with cooling considered if $B_0$ is very small) 
is equal to the observed one for $p=2.4$ and $s \sim 1.7$. Hence, we conclude
that either $p=2$, $s=2$, and $B_0 = 30-50$\,G, or $s<2$, $p\sim 2.4$ and $B_0 \sim 0$\,G,
can explain the radio data for this supernova. 
In table \ref{SelfConsTab} (row 2) we summarize the values of $s$, $p$, and $B_0$
discussed for this supernova.

\subsection{SN\,1993J}

This is the radio supernova with best-observed light curves and expansion 
curve (see P\'erez-Torres, Alberdi \& Marcaide \cite{PerezTorres2002}; 
Bartel et al. \cite{Bartel2002}; Marcaide et al. \cite{Marcaide2009}; 
Weiler et al. \cite{Weiler2007}; Mart\'i-Vidal et al. \cite{MartiVidalI}, 
\cite{MartiVidalII}; and references therein). 

Fitting their observed light curves (taken until $\sim$4900 days after explosion)
Weiler et al. (\cite{Weiler2007}) obtained $\alpha = 0.81$,
$\delta = -1.88$, and $\beta = -0.73$. Therefore, without considering electron 
cooling, from the fitted $\alpha$ we obtain $p = 2.6$ and, from the 
expansion index reported in Mart\'i-Vidal et al. (\cite{MartiVidalII}) at 
late epochs ($m = 0.87$), we obtain $s = 1.6$. Applying now Eq. \ref{betaEq}, 
we derive $\beta = -0.44$, which is inconsistent with the value fitted to the 
light curves ($\beta=-0.73$).

However, if we decrease $m$ down to 0.82, we can obtain a self-consistent solution 
for $\beta$, using Eq. \ref{betaEq}. This seems to be a strong evidence of a CSM 
radial density profile with an index $s<2$.
Also, Mioduszewski, Dwarkadas \& Ball (\cite{Miodus2001})
simulated radio images and the radio light curves of SN\,1993J without taking
radiative cooling into account, and claimed that $s \sim 1.7$ provides the best
fit to the data.

However, the evidence of $s<2$ coming from Eq. \ref{betaEq}, and from the fit of 
the optically-thick part of the radio lightcurves, holds as long as the temperature
of the thermal CSM electrons is taken constant throughout the whole extent
of the CSM (to be able to use $\delta = m\,(1-2s)$), which is not likely to apply
in the case of SN\,1993J (Fransson \& Bj\"ornsson \cite{Fransson1998};
Mart\'i-Vidal et al. \cite{MartiVidalII}). Additionally, more recent analyses of the
X-ray data from SN\,1993J also discard the models with $s<2$ (Nymark, Chandra \& 
Fransson \cite{Nymark2009}; Chandra et al. \cite{Chandra2009}).

From their simultaneous analysis of the complete light curves and expansion curve 
of SN\,1993J, Mart\'i-Vidal et al. (\cite{MartiVidalII}) reported 
$B_0 = 65.1 \pm 1.6$\,G and $p = 2.59 \pm 0.01$ for $s = 2$. It was also noted 
by these authors that using values of $s < 2$ resulted in poor fits to the data. 
From a much time-limited set of flux-density measurements, Fransson \& Bj\"ornsson 
(\cite{Fransson1998}) fitted a similar magnetic field for day 5 after explosion 
($B_0 \sim 68$\,G) using also $s = 2$, although they fitted a different energy 
index for the electron distribution ($p=2.1$).

Which magnetic field do we obtain for SN\,1993J with our approach? 
Opacity effects in the supernova ejecta may affect the spectral 
index and $\beta$ at different frequencies and for different times 
(Marcaide et al. \cite{Marcaide2009}; Mart\'i-Vidal et al. 
\cite{MartiVidalII}). Therefore, 
in our approach we must use the values of $\beta$ and $\alpha$ fitted to 
the subset of data where such ejecta-opacity effects are minimum or non-existent, 
and not those fitted to the whole dataset. Using the 5\,GHz and 8.4\,GHz 
data of Weiler et al. (\cite{Weiler2007}) from day 300 to day 1000 after 
explosion, we obtain $\beta = -0.78\pm 0.05$ at 5\,GHz and $\beta = -0.79\pm 0.08$
at 8.4\,GHz. The average spectral index between 8.4 and 5\,GHz at these 
epochs is $\alpha = 0.98 \pm 0.19$. 

This spectral index implies $p = 3.0$ or lower. For any value of $p$, neither 
$s=1.6$ nor $s=2.4$ yield self-consistent estimates of $B_0$ using our approach.
This is an additional evidence of a CSM with $s = 2$ for SN\,1993J.
Assuming now that $s=2$, we estimate from Eqs. \ref{ApproxBetaEq} and 
\ref{ApproxAlphaEq} that $p\sim2.4$ and $B_0 = 60 - 80$\,G.

In table \ref{SelfConsTab} (row 3) we summarize the values of 
$s$, $p$, and $B_0$ discussed for this supernova.

The range of values of $B_0$ estimated this way is in agreement with the 
estimates reported in Fransson \& Bj\"ornsson (\cite{Fransson1998}) and 
Mart\'i-Vidal et al. (\cite{MartiVidalII}). Nevertheless,
here we have used a subset of the observed light curves, to avoid the undesired
contribution of ejecta-opacity effects in our rough radiative-cooling model.

Fransson \& Bj\"ornsson (\cite{Fransson1998}) and 
Mart\'i-Vidal et al. (\cite{MartiVidalII}) also discussed on the particle-field
energy equipartition, based on their fitted magnetic fields. In both 
papers, it is concluded that, to obtain energy equipartition, an acceleration 
efficiency of the ions similar to that of the electrons 
(i.e., $k \gg 1$ in Eq. \ref{EquipEq}) should take place in the shock.

\subsection{SN\,2008iz}

Marchili et al. (\cite{Marchili2009}) reported a 5\,GHz light curve for
this supernova, taken with the Urumqi telescope. Brunthaler et al. 
(\cite{Brunthaler2010}) reported VLBI observations 
from which the explosion date and the expansion velocity could be estimated. 
Marchili et al. (\cite{Marchili2009}) estimated an equipartition magnetic field
between 0.3\,G and 2.1\,G (for $k=1$ and $k=2000$, respectively) at day 63 after
explosion. Assuming $s=2$ (i.e., $B \propto t^{-1}$), it results in a 
magnetic field between 3.8 and 26.5\,G at day 5 after explosion.

If we use our approach, the spectral index, $\alpha = 1.08 \pm 0.08$ 
(Marchili et al. \cite{Marchili2009}; Brunthaler et al. \cite{Brunthaler2010}), 
is compatible with $p \sim 3$ or lower. However, using $\beta = -1.43 \pm 0.05$ 
(Marchili et al. \cite{Marchili2009}) and $m \sim 0.89$ (derived from a set of 
VLBI observations; Brunthaler et al. \cite{Brunthaler2011} and Brunthaler et al. 
in prep.), we find a self-consistent 
magnetic field of $\sim100$\,G for $p = 2.6$ 
and $s = 2.4$ (see Figs. \ref{fig1b} and \ref{fig2b}), much larger than that 
reported in Marchili et al. (\cite{Marchili2009}).

However, if $p \sim 3$, the magnetic field would be close to 
0\,G, regardless of the value of $s$ (in order to explain the spectral index). 
Now, if we set $s = 2$, we obtain $\beta \sim -1.3$ for $m = 0.89$. This value 
is close to, but lower than, the observed one, and would increase if $s$ would 
be slightly larger than 2. Indeed, the uncertainties in $m$, $\alpha$, and $\beta$ 
can still make possible $s=2$ for $p \sim 3$. In any case, the magnetic field in 
the emitting region can be arbitrarily small if $p \sim 3$, and we cannot favor 
neither this possibility nor the estimate of $B_0 \sim 100$\,G obtained for 
$p = 2.6$.  

In table \ref{SelfConsTab} (row 5) we summarize the values of $s$, $p$, and $B_0$
discussed for this supernova.

\subsection{Other RSNe}

In the cases of RSNe where only the radio light curves are available, it is still 
possible to infer some information on magnetic fields and density structure of the 
CSM and/or ejecta, although with several additional assumptions. In this section, 
we study two cases which we consider interesting compared to other more 
typical RSNe.

\subsubsection{Radio transient in M\,82}

The discovery of a new transient in M\,82 has been recently 
reported in Muxlow et al. (\cite{Muxlow2009}), and a light curve with a practically 
constant flux density has been reported in Muxlow et al. (\cite{Muxlow2010}), with
an spectral index of $\sim 0.7$. Indeed, looking at their Fig. 2, the flux density
at 1.6\,GHz seems to be slightly increasing. If this transient in the starburst 
galaxy M\,82 is a supernova, it would be a so special case, since $\beta \sim 0$. It is 
not possible to obtain such value of $\beta$, unless $s<2$ (see Eq. \ref{betaEq}), since
the highest value of $m$ is 1 and $p$ is assumed to be larger than 1. Indeed, 
from Eq. \ref{betaEq} we obtain $s \sim 0.6$, for $p=2$, and $s \sim 1.3$, for the 
extreme case $p = 1$. Therefore, a plain light curve is a strong evidence of a CSM density 
profile much shallower than the canonical case $s=2$. In any case, another condition 
for $\beta \sim 0$, regardless of the strength of the magnetic field, is that $m \sim 1$ 
(see Fig. \ref{fig1}). Therefore, two clear conclusions
can be extracted for this transient, provided it is a supernova: 1) the index of the CSM 
density profile is $s<2$ and 2) the deceleration index must be $m \sim 1$. Both conclusions
imply that the index of the ejecta density profile, $n$, must be very large ($n = 20$, 
or even higher, see Eq. \ref{mEq}). In regard of the spectral index, from Figs. 
\ref{fig2} and \ref{fig2b} we conclude that the magnetic field would be up to 
$B_0 \sim 20$\,G, assuming $s = 1.6$ and $p = 2$, and lower for larger $p$.

\subsubsection{SN\,2000ft}

Supernova SN\,2000ft was discovered 
by Colina et al. (\cite{Colina2001}). P\'erez-Torres et al. (\cite{Perez2009}) presented 
an eight-year long radio monitoring of this supernova, located in the circumnuclear
starburst of NGC\,7469 (a Luminous Infra-red galaxy, LIRG, at
a distance of 70 Mpc; Sanders et al. \cite{Sanders2003}). 
P\'erez-Torres et al. (\cite{Perez2009}) followed the approach of Weiler et al. 
(\cite{Weiler2002}) to fit the evolution of the radio light curves, using a standard 
value of $s=2$ for the CSM. This analysis resulted in a
value for the spectral index $\alpha=1.27$ and a power-law time
decay index $\beta=-2.02$. 
In addition, they also needed to include a foreground absorber, likely an H II
region, to account for the non-detection of radio emission at
frequencies around and below 1.7 GHz, in agreement with the observations
reported by Alberdi et al. (\cite{Alberdi2006}). 

While the value of $\alpha$ reported for SN\,2000ft is not surprising, the value of 
$\beta$ is much larger (in absolute value) than those typically found in RSNe. From Eq. 
\ref{betaEq}, it is possible to obtain values of $\beta$ similar to that of SN\,2000ft 
if $s > 2$ (see Figs. \ref{fig1} and \ref{fig1b}), although a low value of $m$ (together 
with a large $B_0$) or 
a large value of $p$ is also necessary to simultaneously explain the steep spectrum 
(see Fig. \ref{fig2b}). If $s = 2$, it is also possible to obtain a self-consistency 
between $\alpha$ and $\beta$, provided $B_0 \sim 0$, $p = 3.54$, and $m = 0.75$.

In any case, we find that SN\,2000ft should be a highly decelerated supernova ($m$ between 
0.7 and 0.8), the CSM density index should be $s=2$ or higher, and the energy distribution
of the electrons must be quite steep ($p = 3$ or higher).

\section{Conclusions}
\label{VI}

We have shown the impact of energy losses of relativistic electrons 
in RSNe, and how they affect the flux-density decay rate of the light curves in 
the optically-thin regime for different values of the magnetic fields and for 
different expansion curves.

If the magnetic-field energy density and the acceleration efficiency of the shock 
scale with the shock energy density, which is very likely the case for RSNe, we 
find that there is a tight relation between expansion index, $m$, spectral index,
$\alpha$, and (optically thin) flux-density decay index $\beta$.

This connection between expansion and flux-density evolution in RSNe can 
be used to estimate the magnetic field of observed RSNe ($B_0$ at a reference 
epoch) as well as its evolution with time for an assumed CSM 
radial density profile and energy index, $p$, of the relativistic electrons.

For a number of well observed RSNe (e.g., SN 1993J in M81), self-consistent solutions 
have been found for $B_0$, $m$, $s$, and $p$. A standard CSM density profile (i.e., $s=2$) 
can explain all observations, although evidences of non-standard values of $s$ are found 
for SN\,1986J and SN\,1979C. The index of the relativistic electron population 
takes rather high values ($p = 2.3 - 3.0$) and the range of magnetic fields 
between all cases is large ($B_0 \sim 20-100$\,G). These large magnetic fields 
imply effective amplification mechanisms in the radio-emitting region, possibly related
to plasma turbulence (see, e.g., Gull \cite{Gull} or Jun \& Norman \cite{Jun}, and 
references therein).
 
Previous analyses of the radio light curves and expansion curves of these RSNe did not
take into account the correct coupling between $m$, $\beta$, and $\alpha$ for different 
magnetic fields. Some of the results previously reported for these supernovae
could, therefore, be internally inconsistent.

The magnetic fields obtained with our approach are in similar to the 
equipartition magnetic fields. For SN\,1979C and SN\,1986J, we obtain a range 
of self-consistent magnetic fields similar to those derived 
from equipartition with a lower acceleration efficiency for ions (i.e., 
low-to-intermediate values of $k$ in Eq. \ref{EquipEq}). Additionally, for 
SN\,1986J there is evidence of $s < 2$, provided the magnetic field is small. 
For SN\,2008iz, either a very low magnetic field (with $s\sim 2$) {\em or} an 
extremely large magnetic field (with $s > 2$) are necessary to model the light 
curve, given the large flux-density decay rate ($\beta =-1.43$). For SN\,1993J, 
we obtain a magnetic field similar to that reported in Fransson \& Bj\"ornsson 
(\cite{Fransson1998}) and Mart\'i-Vidal et al. (\cite{MartiVidalII}), although 
we use in our approach a subset of flux-density observations (and not the whole
data set), to avoid possible biasing effects coming from the ejecta opacity 
(Mart\'i-Vidal et al. \cite{MartiVidalII}).

For the RSNe that will be detected in the future (the large sensitivity of the 
forthcoming radio observatories, like ALMA and SKA, will allow the detection and 
monitoring of many other RSNe), it will be necessary, in light of the results here
reported, to study the connection between their expansion and flux-density evolution,
in order to obtain self-consistent results for the CSM profile, the electron energy 
index, and the magnetic field, based on the observed spectral index, expansion curve,
and flux-density decay index.

\begin{acknowledgements}

IMV is a fellow of the Alexander von Humboldt foundation in Germany.
MAPT acknowledges support by the Spanish
Ministry of Education and Science (MEC) through grant AYA
2006-14986-C02-01, and by the Consejer\'{\i}a de Innovaci\'on,
Ciencia y Empresa of Junta de Andaluc\'{\i}a through grants
FQM-1747 and TIC-126.

\end{acknowledgements}

\onecolumn

\begin{appendix}

\section{Derivation of the source function of relativistic electrons}
\label{SourceApp}

The source function is related to the acceleration of part of the electrons 
from the CSM,
as they interact with the expanding supernova shock. If we assume that the 
acceleration efficiency scales with the energy density of the 
shock (see Fransson \& Bj\"ornsson \cite{Fransson1998} for a discussion of
different possibilities and how they fit to the observations of SN\,1993J),
the density of electrons instantaneously accelerated at a given time, $t$, will be

\begin{equation}
n_{\mathrm{rel}} \propto n_{\mathrm{cs}}\,V^2,
\label{relatdensApp}
\end{equation}

\noindent where $V$ is the velocity of the expanding supernova shock and $n_{\mathrm{cs}}$ 
is the density of the recently-shocked CSM, both quantities computed for the same time $t$. 
Since the radius of the shock is $r \propto t^m$, the number of relativistic electrons 
injected between $r$ and $r+dr$ is

\begin{equation}
N_{\mathrm{rel}} = F_{\mathrm{rel}}\,n^0_{\mathrm{cs}}\,\left(\frac{r}{r_0}\right)^{-s}\,V^2\,4\,\pi\,r^2\,dr,
\label{TotNrelApp}
\end {equation}

\noindent where $s$ is the index of the CSM radial density profile, $n^0_{\mathrm{cs}}$
and $r_0$ are the CSM density and shock radius at a given reference epoch, $t_0$, and
$F_{\mathrm{rel}}$ is the acceleration efficiency (or fraction of CSM electrons that 
are accelerated) at the same epoch $t_0$. 

These electrons are distributed according 
to $N \propto E^{-p}$ (with $E$ running from $E_{\mathrm{min}} = m_e\,c^2$ to infinity, 
being $m_e$ the electron mass). Therefore, the conservation of the number of electrons 
implies

\begin{equation}
N_{\mathrm{rel}} = K \int_{m\,c^2}^{\infty}{E^{-p}\,dE}.
\label{NormFac1}
\end{equation}

\noindent The factor $K$ accounts for the normalization of the electron energy
distribution. This factor is just

\begin{equation}
K = \frac{p-1}{E_{\mathrm{min}}^{1-p}}.
\label{NormFac2}
\end{equation}

Hence, the source function (i.e., the energy distribution of electrons shocked 
between $r$ and $r+dr$) is

\begin{equation}
S(E,r) = F_{\mathrm{rel}}\,N^0_{\mathrm{cs}}\frac{p-1}{E_{\mathrm{min}}^{1-p}}\left(\frac{r}{r_0}\right)^{5-s-3/m}\,E^{-p},
\label{SourceEqApp}
\end{equation}

\noindent which, in terms of time (given that $r/r_0 = (t/t_0)^m$), reduces to Eq. \ref{SourceEq}.

\section{Radiative and adiabatic energy loss vs. free-free loss in RSNe}
\label{RadAdiFreeApp}

The rate of energy loss due to synchrotron radiation and adiabatic expansion, 
$\dot{E}_{\mathrm{r}}$ and $\dot{E}_{\mathrm{a}}$, are given in Eq. \ref{DotEEq1} 
and \ref{DotEEq2}, respectively. In 
regard of the energy loss due to free-free interactions with the CSM, we have 
(Pacholczyk \cite{Pacholczyk1970})

\begin{equation}
\dot{E}_{\mathrm{f}} \sim f_1\,n_{\mathrm{cs}}\,E,
\label{LossFreeEq}
\end{equation}

\noindent where $f_1 \sim (1-8) \times 10^{-16}$ in cgs units. The exact value depends 
on the level of ionization of the nuclei in the CSM (lower values of $f_1$ correspond to 
higher levels of ionization, which are expected in the shocked CSM)\footnote{There is an 
additional contribution to $\dot{E}_{\mathrm{f}}$ that can be written as a modifying factor
of $f_1$, which depends on $\log{E}$. We have neglected this small correction.}. 
Since an electron with energy $E$ in interaction with a magnetic field $B$ emits 
synchrotron radiation mostly at its critical frequency (given by 
$\nu \sim c_1\,B\,E^2$, where $c_1 = 6.27\times10^{18}$ in cgs units, 
Pacholczyk \cite{Pacholczyk1970}), the ratio of radiative loss to free-free loss 
for electrons emitting at the critical frequency $\nu$ is

\begin{equation}
\frac{\dot{E}_{\mathrm{r}}}{\dot{E}_{\mathrm{f}}} = \frac{c_2}{f_1}\sqrt{\frac{\nu}{c_1}}\,
\frac{B_0^{3/2}}{n_0}\left(\frac{t}{t_0}\right)^{(m(6-s)-6)/4}.
\label{RatioRad}
\end{equation}

\noindent where we have used the time evolution of $B$ given in Eq. \ref{magEq}. On the other
hand, the ratio of diabatic to free-free energy loss for electrons emiting at the same 
frequency is

\begin{equation}
\frac{\dot{E}_{\mathrm{a}}}{\dot{E}_{\mathrm{f}}} = \frac{m}{f_1\,n_0\,t_0^{m\,s}}\,t^{m\,s-1}.
\label{RatioAdi}
\end{equation}

The ratios in Eqs. \ref{RatioRad} and \ref{RatioAdi} evolve as power laws of time, whose indices depend on 
$s$ and $m$. Therefore, for some combinations of $s$ and $m$, the ratios will grow with time (and 
radiative and adiabatic losses will dominate over free-free losses), but for other combinations the 
ratios will decrease with time, and free-free losses may be comparable to the other contributions 
at late times.

In the case of radiative vs. free-free losses, the time index in Eq. \ref{RatioRad} is positive if

\begin{equation}
m > \frac{6}{6+s},
\label{mAd}
\end{equation}

\noindent which implies values of $m$ larger than 0.75 for the canonical case $s=2$, although 
slightly higher values of $m$ for lower $s$ (for instance, $m>0.78$ if $s = 1.6$). However, even 
if $m$ is lower than these values (so the ratio $\dot{E}_{\mathrm{r}}/\dot{E}_{\mathrm{f}}$ decreases 
with time), radiative losses will still be 
higher than free-free losses for the times and observing frequencies of interest. For instance,
with an initial magnetic field as low as $B_0 = 10$\,G (at day 5 after explosion!) and a CSM 
density as large as $n_0 = 10^9$\,cm$^{-3}$, the ratio is $\sim 15$ at day 1000 after explosion, 
for $m = 0.70$, observing at 5\,GHz. 

In the case of the ratio of adiabatic losses to free-free losses, it increases with time if 
$m>1/s$. This relation allows expansion indices as low as 0.63 if $s = 1.6$, and even lower 
values for higher $s$. Even in the (so special) cases where we would find $m<1/s$, adiabatic 
losses would still be higher than free-free losses for the times and observing frequencies of 
interest. For instance, if $m=0.59$ (i.e., the minimum possible value compatible with the 
Chevalier model for $s = 1.6$) and $n_0$ is as large as 
$10^9$\,cm$^{-3}$, this ratio is $\sim 10$ at day 1000 after explosion observing at 5\,GHz. 

In any case, for strongly decelerated RSNe (i.e., with low $m$) free-free losses might not be 
completely negligible, depending also on the CSM density (higher density implies larger free-free 
energy losses). In these special cases, the connection 
between $\beta$ and $m$ may not only depend on $s$, $p$, and $B_0$, but also on $n_0$. Hence,
Eq. \ref{contEq} will have to be individually integrated for each of these cases.

\section{Integral solution of the continuity equation in energy space}
\label{AppA}

The energy loss of an electron with energy $E$ at time $t$ is given in Eq. \ref{DotEEq}.
Here we rewrite the equation,

\begin{equation}
\dot{E} = -c_2\frac{2}{3}B_{\mathrm{ref}}^2\left(\frac{t}{t_{\mathrm{ref}}}\right)^{\rho}E^2 -m\frac{E}{t},
\label{EdotApp}
\end{equation}

\noindent where $\rho = 2(m-s)-2$ and $B_{\mathrm{ref}}$ is the magnetic field at a reference epoch 
$t_{\mathrm{ref}}$ (we do not use the subindex $0$, to distinguish the reference epoch from the initial time 
of integration, $t_0$, see below). The solution to this equation is

\begin{equation}
E(t) = \frac{(1+\rho-m)t_0^m E_0}{(1+\rho-m)t^m+2/3\,c_2B_{\mathrm{ref}}^2 E_0 t_{\mathrm{ref}}^{-\rho}(t^{1+\rho}t_0^m-t_0^{1+\rho}t^m)},
\label{ESolApp}
\end{equation}

\noindent where we have computed the integration constant by assuming $E=E_0$ at 
$t = t_0$ (i.e., the initial time of integration). We now write $E_0$ in terms of $E$:

\begin{equation}
E_0 = \frac{(m-1-\rho) E}{(m-1-\rho)(t_0/t)^m+2/3\,c_2B_{\mathrm{ref}}^2 E t_{\mathrm{ref}}^{-\rho} (t^{1+\rho}(t_0/t)^m-t_0^{1+\rho})}.
\label{E0SolApp}
\end{equation}

At a given time, $t_0$, the electrons being shocked (let us call them $N^0(E_0)$) are distributed 
as $S(t_0)E_0^{-p}$ (where $S(t_0)$ is given in Eq. \ref{SourceEq}). Since the number of 
electrons is conserved, at a later time, $t$, these electrons (i.e., not all the electrons, but
just those shocked at time $t_0$) will have the distribution $N^0(E)$ given by

\begin{equation}
N^0(E)dE = N^0(E_0)dE_0 \rightarrow N^0(E) = N^0(E_0)\frac{dE_0}{dE}.
\label{NdEEqApp}
\end{equation}

\noindent Applying Eq. \ref{E0SolApp}, and its derivative, to Eq. \ref{NdEEqApp}, we obtain

\begin{equation}
N^\tau(E) = \frac{S(\tau)(m-1-\rho)^{2-p}E^{-p}(\tau/t)^m}
{\left((m-1-\rho)(\tau/t)^m+2/3\,c_2B_{\mathrm{ref}}^2 E t_{\mathrm{ref}}^{-\rho}((\tau/t)^m\,t^{\rho+1}-\tau^{1+\rho})\right)^{2-p}},
\label{TauTApp}
\end{equation}

\noindent where $t_0$ has been replaced by $\tau$, which can take any value in the evolution time 
of the supernova.
A similar approach was described in Pacholczyk (\cite{Pacholczyk1970}) (see his Sect. 6.3), although 
a constant and 
homogeneous magnetic field was used. We notice that Eq. \ref{TauTApp} is physically meaningful only
when the power in the denominator is that of a positive number. Let us call $t_F$ the value of $\tau$ such 
that the denominator of Eq. \ref{TauTApp} vanishes. For a given supernova, this value depends on $E$ and 
$t$. For larger $t$, all the electrons shocked at time $\tau$ will have energies below $E$. In other words, 
no electrons shocked at time $t_F$ (and earlier times) contribute to the electron distribution at 
time $t$ for energies above $E$. 

It is now straightwforward to conclude that the total number of electrons at time $t$ and 
energy $E$ will just be the addition of all the (evolving) contributions of the source function between 
the beginning of the expansion, $t_0$, and $t$ (these contributions are given by Eq. \ref{TauTApp}). The 
resulting integral is

\begin{equation}
N(E,t) = \int_{\,t_F}^{\,t}{\frac{S(\tau)(m-1-\rho)^{2-p}E^{-p}(\tau/t)^m\,d\tau}
{\left((m-1-\rho)(\tau/t)^m+2/3\,c_2B_{\mathrm{ref}}^2 E t_{\mathrm{ref}}^{-\rho}((\tau/t)^m\,t^{\rho+1}-\tau^{1+\rho})\right)^{2-p}}}.
\label{ExactSol}
\end{equation}

\section{Why are the light curves flatter if we consider radiative cooling?}
\label{betaradApp}

From Eq. \ref{IBN}, we can approximate the value of $\beta$ by assuming that, at 
any time, the electron energy distribution does not differ so much from the 
canonical one, $N \propto E^p$. Under such an assumption (valid for a rough qualitative 
discussion), it can be shown that

\begin{equation}
\beta = \frac{d\,\mathrm{ln}(I)}{d\,\mathrm{ln}(t)} \propto \frac{\dot{I}}{I} = \frac{1+p}{2}\frac{\dot{B}}{B} + \frac{\dot{N}}{N}.
\label{betaAppr}
\end{equation}

The term with $\dot{B}/B$, which is negative and is related to the different electron emissivities 
under different magnetic fields, has the same effects on $\beta$ 
either if radiative cooling is considered or not. In regard of the term $\dot{N}/N$, which is 
positive and is related to the different total emissivity for different number of electrons, 
when radiative cooling is {\em not} considered it takes the form

$$ \frac{\dot{N}(E,t)}{N(E,t)} = \frac{S(E,t)}{\int_{t_0}^{t}{S(E,t')\,dt'}}, $$

\noindent where $S(E,t)$ is the source function given in Eq. \ref{SourceEq}. However, 
if radiative cooling {\em is} considered, we have instead

$$ \left(\frac{\dot{N}(E,t)}{N(E,t)}\right)_{\mathrm{Cool}} = \frac{S(E,t)-c(E,t)}{\int_{t_0}^{t}{(S(E,t')-c(E,t'))\,dt'}}, $$

\noindent where $c(E,t) = -\nabla_E(N\dot{E})$ and $\dot{E}$ is given in Eq. \ref{DotEEq}.
The function $c(E,t)$, which indirectly depends on $N$, approaches 0 as time goes by. Therefore, 
the numerator, $\dot{N}(E,t)$, if cooling is considered, will tend to that of the cooling-free case, 
while the denominator, $N(E,t)$, will always be smaller in the case with cooling. Hence,

\begin{equation}
\left(\frac{\dot{N}(E,t)}{N(E,t)}\right)_{\mathrm{Cool}} > \frac{\dot{N}(E,t)}{N(E,t)}.
\label{IneqApp}
\end{equation}

From this equation, it is easy to see that the positive contribution to $\beta$ provided by 
$\dot{N}/N$ (see Eq. \ref{betaAppr}) is larger if radiative cooling is considered. The 
corresponding value of $\beta$ is, therefore, closer to 0 (the light curve is flatter). 

The inequality in Eq. \ref{IneqApp} may have important observational effects until very late 
times (several years), when we will have

$$\int_{t_0}^{t}{S(E,t')\,dt'} >> \int_{t_0}^{t}{c(E,t')\,dt'}.$$

Even then, the {\em average} value of $\beta$ in the whole optically-thin part of the light curve
may still differ from that of the cooling-free case.

\end{appendix}

\end{document}